# Water permeation pathways in laminated organic single-crystal devices

Ryo Nouchi,[1,2,a)] Yoshiaki Ishihara,[1] and Susumu Ikeda[3]

[1]*Department of Physics and Electronics, Osaka Prefecture University, Sakai 599-8570, Japan*

[2]*PRESTO, Japan Science and Technology Agency, Kawaguchi 332-0012, Japan*

[3]*WPI-Advanced Institute for Materials Research (WPI-AIMR), Tohoku University, Sendai 980-8577, Japan*

[a)] Author to whom correspondence should be addressed: r-nouchi@pe.osakafu-u.ac.jp

**ABSTRACT**

Water permeation pathways in electronic devices should be eliminated for the suppression of operational instabilities. We investigated possible pathways in field-effect transistors based on a laminated single crystal (SC) of an organic semiconductor, rubrene. Water-induced instabilities were found to be more obvious with a thicker rubrene SC. Furthermore, molecular dynamics calculations of water diffusion on a rubrene SC showed that no water molecules penetrated the SC under our simulation conditions. These findings indicate that a space at the SC/substrate interface is a dominant pathway. The present study clearly shows the importance of conformality of SC lamination onto the underlying substrate.

# INTRODUCTION

Stable operation in air is an indispensable requirement for most electronic devices. Instabilities are often induced by adsorption and diffusion of air molecules, such as oxygen and water. The effect of the former molecule is rather simple: it significantly degrades the device characteristics if easily oxidizable components are used in the device structure. The latter molecule possesses a permanent electric dipole, and thus, various mechanisms have been proposed to explain the effect of water molecules on the operation of electronic devices, such as field-effect transistors (FETs). The mechanisms include diffusion of water-related species into the gate insulator of FETs,[1-3] water-induced polarization of the gate insulator,[4,5] and charge carrier trapping at or near the channel formed at the semiconductor/insulator interface;[6-10] all of which are related to the gate insulator and the interface with the gate insulator. These gate-related phenomena are induced after diffusion and penetration of water into the semiconductor layer, and hence, they should be significant in loosely packed semiconductors.

Organic semiconductors are now regarded as a channel material that is promising for constructing large-area, flexible electronic devices.[11-13] However, inter-molecular interactions are mainly attributed to weak van der Waals forces, from which it is expected that penetration of small molecules from the air into the semiconductor layer occurs rather easily. Therefore, the effect of the water-related instabilities is expected to be significant in organic electronic devices. In the case of organic FETs (OFETs), various mechanisms of their operational instabilities have been extensively studied. The mechanisms have been mainly attributed to the gate insulator and the interface with the gate insulator, as exemplified above.[14] One of the authors (Nouchi) recently proposed a mechanism related to the source/drain contacts:[15] water molecules reach the source/drain electrode surfaces and induce a change in electrode work function in response to the applied drain bias, which was evidenced by switching from symmetric to rectified current–voltage characteristics in two-terminal devices (even devices with symmetrically fabricated electrodes). This electrode-related mechanism also requires permeation of water molecules into the device structure.

In order to suppress such water-induced instabilities, water permeation pathways in OFETs should be determined for proper elimination of the instabilities. It is generally accepted that major pathways for water permeation through any substance are defects or pinholes within the substance.[16-18] Such defects are universally present in polycrystalline films as grain boundaries. Thus, it is natural to say that single crystalline films are a superior barrier against water permeation than their polycrystalline counterparts. Since grain boundaries act as a scattering/trapping center for charge carrier transport,[19-21] single crystals (SCs) are also desired from the viewpoint of OFET performance. In this study, we fabricated bottom-contact, bottom-gate FETs with a SC of an archetypal organic semiconductor, rubrene, as schematically shown in Fig. 1(a). Thus, among various permeation pathways depicted in Fig. 1(b), defects and grain boundaries can be excluded as the possible pathway. In this device configuration, SCs are usually laminated by manually picking up a SC and placing it onto a target substrate with pre-defined electrodes, which can avoid formation of ill-defined contacts, as is problematic in top-contact devices. The interaction between the SC and the underlying substrate is categorized as a weak van der Waals interaction, which is the same as the inter-molecular interaction within the SC. Thus, two possible pathways for water permeation would be inter-molecular spacings in the SC and a space between the SC and the underlying substrate. We characterized SCFETs with different SC thicknesses and found that the SC/substrate interface is a dominant pathway for water permeation into laminated SC devices. Molecular dynamics (MD) simulations were also conducted on the diffusion of water molecules placed near a rubrene SC and water permeation through the inter-molecular spacings was found to be unlikely to occur. These findings indicate the importance of conformality of the SC lamination. This information should be general knowledge for analyses of the device characteristics based on semiconductor channels laminated by weak van der Waals forces.

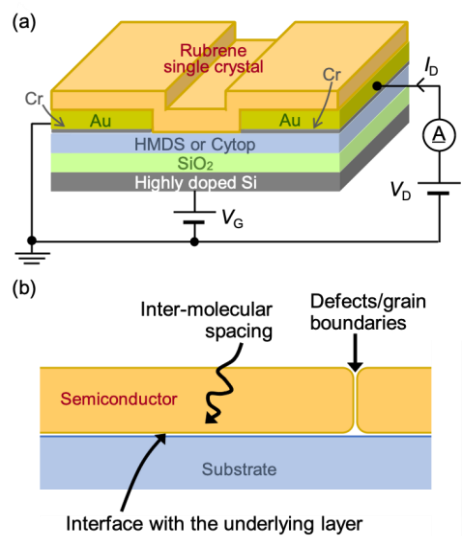

**FIG. 1.** (a) Schematic diagram of fabricated rubrene SCFETs. (b) Possible pathways for water permeation in FET configurations.

**EXPERIMENTAL**

A highly doped Si wafer with a 300 nm-thick thermal oxide layer was used as a substrate. The highly doped Si wafer and the thermal oxide layer serve as the gate electrode and the gate dielectric, respectively. The substrate was cleaned with acetone and 2-propanol using an ultrasonic bath, followed by an oxygen plasma treatment. The surface OH groups[22] on the cleaned substrate were then passivated by the treatment with 1,1,1,3,3,3-hexamethyldisilazane (HMDS; >96%, Wako) or by coating with a fluoropolymer layer (Cytop™ CTL-809M, AGC).[23] For the HMDS treatment, the cleaned Si substrate was immersed overnight in a hexane solution of HMDS, with the volume ratio of HMDS to hexane of 1/9. After removal from the solution, the substrate was cleaned by pure hexane in an ultrasonic bath in order to detach physisorbed multilayers of HMDS molecules. A surface profiler (Dektak150, Veeco) was used to determine the typical thickness of the fluoropolymer coating to be ca. 100 nm. After the passivation procedure, Au electrodes with a thickness of 15 nm were formed by thermal evaporation together with a Cr adhesion layer. The electrodes were patterned by a metal shadow mask to form a 50 μm-long channel. For the semiconductor layer, a rubrene SC was synthesized by the physical vapor transport (PVT) of the source powder (≥ 98%, Sigma-Aldrich). The growth process was repeated three times in order to improve the purity of the crystal. Finally, the single crystal was manually laminated onto the pre-defined Au electrodes.[24,25] A molecularly flat surface of the rubrene crystals was confirmed by an atomic force microscope investigation (Hitachi High-Technologies, AFM5200S), from which we infer that the rubrene crystals are single-crystalline (not polycrystalline). An optical micrograph of a FET fabricated in this study is shown in Figure 2(a). Electrical measurements were conducted using a semiconductor device analyzer (B1500A, Keysight) in ambient air at room temperature.

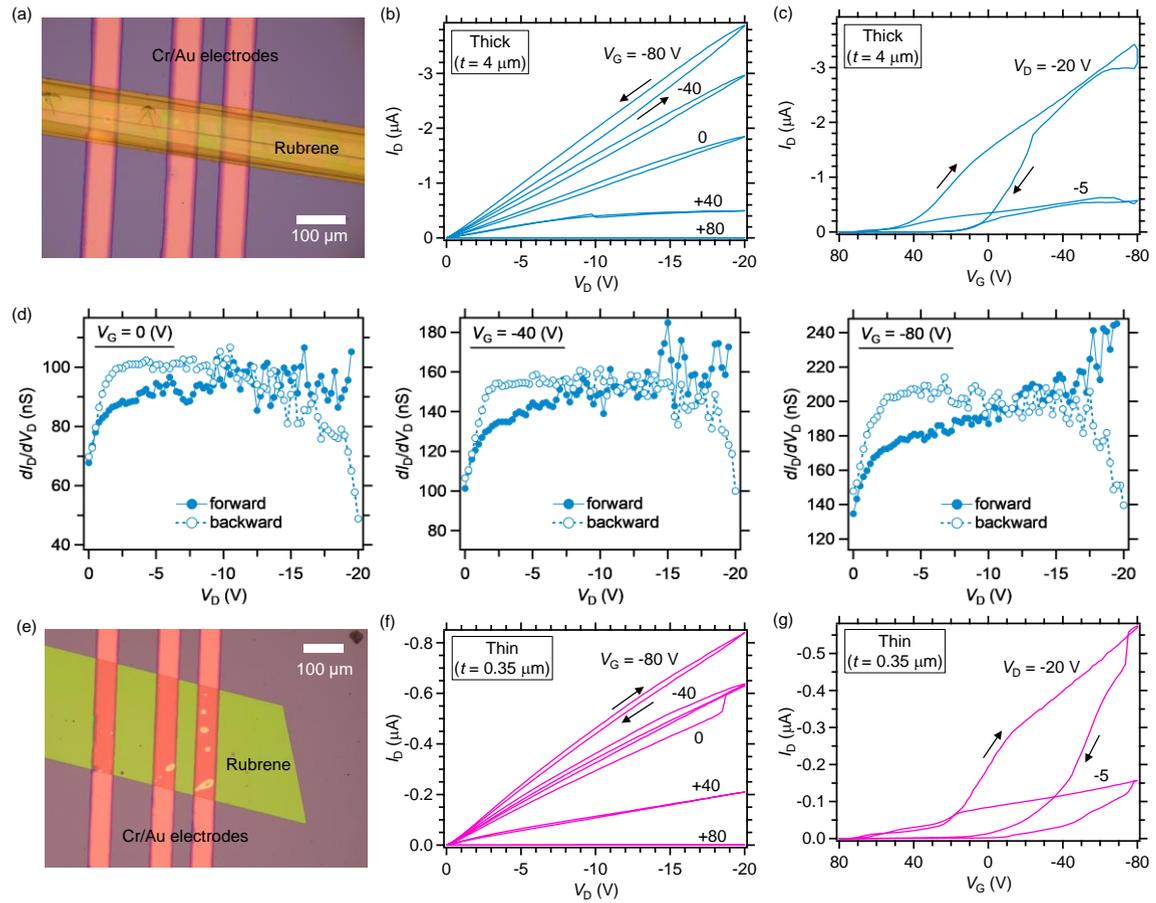

**Fig. 2.** Rubrene SCFETs based on (a–d) a thick and (e–g) thin SC. The substrate surface was passivated with HMDS. (a,e) Optical micrograph. (b,e) Output characteristics. (c,f) Transfer characteristics. (d) $V_D$ dependence of the differential conductance of the SCFET with a thick SC.

## RESULTS AND DISCUSSION

Figures 2(b) and 2(c) show output ($I_D$–$V_D$) and transfer ($I_D$–$V_G$) characteristics of a rubrene SCFET with a crystal thickness of 4 μm, where $I_D$ is the drain current, $V_D$ is the drain voltage, and $V_G$ is the gate voltage. The output characteristics show a counterclockwise-type hysteresis, where the magnitude of the $I_D$ in the backward sweep is higher than in the forward sweep of the bias voltage. This indicates that the SCFET has instabilities that accompany an increase in $|I_D|$. In our previous study based on short-channel two-terminal devices with an Au–rubrene–Au configuration, an electrical current was found to increase only in the presence of water when an external voltage was applied to the device.[15] The increase in electrical current was explained by an increase in the work function of the anode, which lowered the Fermi level and enhanced hole injection into the valence band derived by the highest occupied molecular orbital (HOMO) of rubrene.[15] Therefore, the counterclockwise-type hysteresis observed in Fig. 2(a) is ascribable to the electrode-related instability that can occur only in the presence of water.

The electrode-related instability discussed here has been understood as a change in the charge-injection barrier height at the electrode contact.[15] The barrier height is known to affect the shape of the low-voltage region in the output characteristics of FETs. A low barrier results in a linear $I_D$–$V_D$ curve in the low-$V_D$ region, and a differential conductance ($dI_D/dV_D$) of this region becomes constant in an ideal case. On the other hand, a high barrier leads to a super-linear curve that is generally explained by thermionic injection, and a differential conductance exhibits an increase as $V_D$ increases. Figure 2(d) shows the $V_D$ dependence of the differential conductance of the device shown in Figs. 2(a-c). As clearly shown in these figures, the differential conductance shows an increasing trend during the forward sweep of $V_D$, and a linear portion is widened during the backward sweep. These results suggest a lower injection barrier during the backward sweep than the forward one, which supports a manifestation of the electrode-related instability in these devices.

To the contrary, the transfer characteristics in Figure 2(c) show a commonly observed clockwise-type hysteresis, where the magnitude of the $I_D$ in the backward sweep is lower than in the forward sweep of the bias voltage. This indicates that the SCFET has instabilities that accompany a decrease in $|I_D|$, which is opposite to what was observed in the output characteristics. This type of hysteresis is typical of most of gate-related instabilities. The discrepancy in the type of observed hysteresis between output and transfer characteristics has been reported in OFETs with source/drain electrodes whose surfaces were modified with low-density self-assembled monolayers of thiol molecules.[26] For the measurement of output characteristics, first $V_G$ is set and then $V_D$ is swept. The $|I_D|$ decrease caused by the gate-related instabilities is expected to mostly finish during the forward sweep of $V_D$. Thus, in the backward sweep, the $|I_D|$ increase by the electrode-related instabilities can be larger than the $|I_D|$ decrease by the gate-related ones. Since the observed hysteresis can be understood as a summation of the electrode and gate-related instabilities, in this case, the counterclockwise-type hysteresis is observed. Therefore, instabilities caused by application of $V_D$, i.e., electrode-related instabilities, dominate the output characteristics, Fig. 2(b), and the opposite holds for the transfer characteristics, Fig. 2(c).

Among various gate-related instabilities, slow polarization of water molecules embedded in the gate dielectric can also cause a counterclockwise-type hysteresis.[27] Although the slow polarization mechanism is one of the gate-related instabilities, a counterclockwise-type hysteresis was observed only in output characteristics. Therefore, we concluded that the slow polarization mechanism was not significant in our devices.

We discuss here which instabilities, the $|I_D|$ increase or decrease, should be employed as a probe of water permeation into the device. The electrode-related $|I_D|$ increase in two-terminal Au–rubrene–Au devices has been confirmed to be detected only in the presence of water in the surrounding environment.[15] On the other hand, a variety of mechanisms have been developed for the gate-related $|I_D|$ decrease, some of which are indeed known to be enhanced by water molecules. However, mechanisms that are not related to water (e.g., trap formation by the bias stress, etc.)[28] should also contribute to the $|I_D|$ decrease. Furthermore, the hydrophobic nature of the passivated surfaces significantly reduces the water-related instabilities at the interface between the gate insulator and the semiconductor layer. From the considerations above, we concluded that the $|I_D|$-increasing hysteresis caused by the electrode-related instabilities could be used as a probe of the water permeation into the device structure.

A water permeation rate via inter-molecular spacings is dependent on the thickness of SCs. Thus, a SC much thinner than that used in Fig. 2(a) was used to fabricate another SCFET as shown in Fig. 2(e). Figures 2(f) and 2(g) show output and transfer characteristics of a SCFET with a crystal thickness of 0.35 μm. The hysteresis observed in both the characteristics is a clockwise type, which indicates that gate-related instabilities dominate them. It should be noted that the hysteresis window in Fig. 2(g) is significantly wider than that in Fig. 2(c). Since the observed hysteresis is a summation of the electrode and gate-related instabilities, even the clockwise-type hysteresis, which is dominated by the gate-related instabilities, is contributed to by the electrode-induced instabilities. Therefore, the wider window of the $|I_D|$-decreasing (clockwise-type) hysteresis in Fig. 2(g) is understood by a smaller contribution from the $|I_D|$-increasing instabilities caused by the electrode-related mechanism. These experimental facts indicate that water permeation becomes more inefficient in a thinner SC. The lower permeation efficiency with the thinner SC indicates that inter-molecular spacings are not a dominant pathway for water permeation.

It should be noted that the FET characteristics shown in Fig. 2 display accidental $I_D$ jumps. This anomaly is a sign of the presence of an incompletely laminated portion within the SC channel. Accidental attachment/detachment of the SC to/from the substrate might be induced by the gate electric field, leading to the current jumps. We frequently experienced that the lamination process became difficult after passivating the $SiO_2$ surface with HMDS.

This difficulty was not encountered in the case of the fluoropolymer coating. These observations might be ascribable to the stronger electrostatic interaction of the rubrene SC with polar surfaces, namely, hydroxyl groups on the bare $SiO_2$ and the fluorinated skeleton of the fluoropolymer. Even if the accidental jumps are taken into consideration, the overall shapes of the transfer and output characteristics in Fig. 2 can be interpreted as the same as the discussions given in the previous paragraphs.

The lamination of a SC is known to become more conformal to the underlying surface as the SC becomes thinner, which is ascribable to the enhanced flexibility of thinner crystals.[29] Thus, the SC with a thickness of 4 μm in Figs. 2(a–d) is more rigid than that with a thickness of 0.35 μm in Figs. 2(e–g). The incompleteness of the lamination provides another permeation pathway, i.e., a space between the SC and the underlying substrate. The importance of the SC/substrate interface is further corroborated by comparing FET characteristics with and without bubbles unintentionally formed during the SC lamination process. Figure 3(a) shows an optical micrograph of a SCFET with such bubbles. Its output and transfer characteristics are shown in Figs. 3(b) and 3(c), which display a counterclockwise and clockwise-type hysteresis, respectively. A SCFET with a similar SC thickness but without such bubbles, Fig. 3(d), was also tested. Its output and transfer characteristics in Figs. 3(e) and 3(f) display the same hysteresis type as those in Figs. 3(b) and 3(c), respectively. However, the SCFET without the bubbles show a smaller counterclockwise-type hysteresis in the output characteristics and a larger clockwise-type hysteresis in the transfer characteristics than that with the bubbles. Thus, the $|I_D|$-increasing type hysteresis caused by the electrode-related instabilities is more significant in the SCFET with the bubbles than in that without the bubbles. This fact also indicates the importance of the SC/substrate interface as a water permeation pathway.

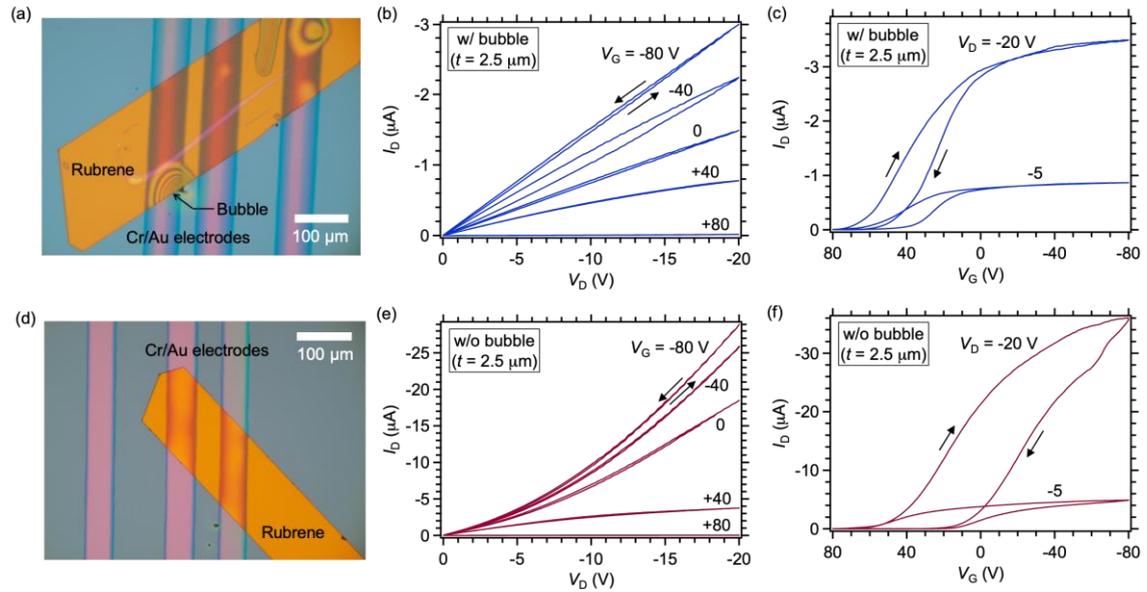

**Fig. 3.** Rubrene SCFETs (a–c) with and (d–f) without bubbles unintentionally formed during the SC lamination process. Since the SC thicknesses were almost the same with these two devices, a difference in electrical characteristics was expected to be small. Thus, the substrate surface was passivated with a fluoropolymer coating in order to strongly suppress the gate-related instabilities that mask the electrode-related instabilities. (a,d) Optical micrograph. (b,e) Output characteristics. (c,f) Transfer characteristics.

Figure 4 shows a relative humidity (RH) dependence of the $I_D$–$V_D$ characteristics of a similar device fabricated by laminating a rubrene single crystal on a bare $SiO_2$ surface with pre-defined Au electrodes. The thickness of the electrodes was 15 nm; the channel length was 30 μm; the thickness of the crystal was 0.69 μm. A low $V_G$ value was selected to minimize the gate-bias stress ($V_G = -8$ V). The vertical axis of Fig. 4(a) is the as-measured $I_D$; that of Fig. 4(b) is the ratio of $I_D$ during the backward ($I_{Db}$) and forward sweep ($I_{Df}$) of $V_D$. The $I_D$-$V_D$ characteristics show a clockwise-type hysteresis, leading to $I_{Db}/I_{Df}$ being less than 1. This result is attributable to a significant gate-bias stress exerted by a large amount of hydroxyl groups on the bare $SiO_2$ surface. Nevertheless, although its behavior is complicated and not monotonic in the low-$V_D$ region, the general trend is that a higher RH value led to a higher $I_{Db}/I_{Df}$ ratio. Since the observed hysteresis is a summation of the current-decreasing and current-increasing hystereses, the fact that the $I_{Db}/I_{Df}$ ratio becomes higher for the higher RH is the confirmation that the current-increasing hysteresis becomes more dominant for higher RH.

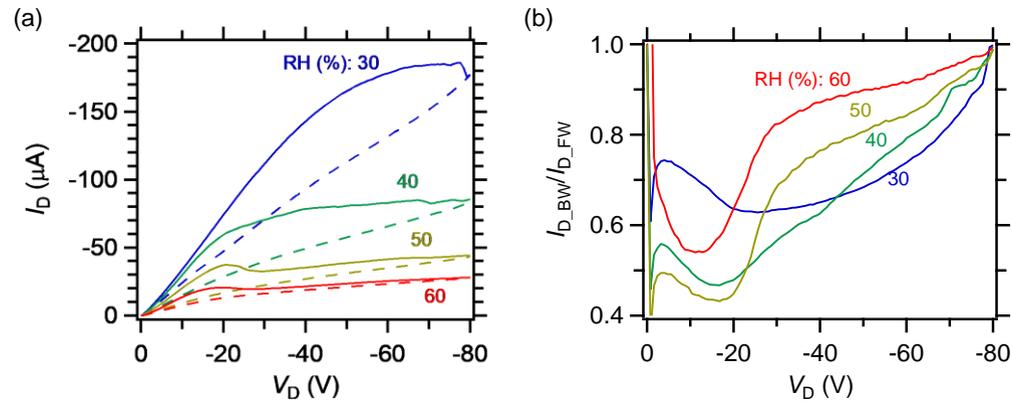

**Fig. 4.** Relative humidity (RH) dependence of the hysteresis in the output characteristics of an SCFET with a rubrene SC laminated on a bare $SiO_2$ surface. (a) $I_D$–$V_D$ characteristics. The solid and dashed lines correspond to the forward and backward sweeps of $V_D$, respectively. (b) Ratio of $I_D$ during the backward ($I_{D\_BW}$) and forward sweep ($I_{D\_FW}$) of $V_D$.

In order to provide further confirmation, we conducted MD simulations on the diffusion of water molecules placed near a rubrene SC. To mimic water permeation towards the electrode/SC interface, water molecules were placed on a rubrene/Au(100)/amorphous-$SiO_2$ stack. Prior to the MD simulations, the partial charge of rubrene and water molecules was determined by Gaussian 09 with the Hartree-Fock method using 6-31G(d) basis sets.

The same system was utilized for the structural optimization of a water molecule. For the structure of a rubrene molecule, we employed a low-temperature structure reported with a rubrene SC grown by the PVT method,[30] and the structure of the molecule was fixed during the calculation of the partial charge carried out by Gaussian 09. The determined molecular structures, including the values of the partial charge, were kept during the MD simulations (rigid model). The full details of the MD simulations are explained elsewhere,[31,32] however, briefly, the water–water interaction was calculated based on the TIP5P potential,[33] and the intermolecular and molecule–substrate interactions were described by the Lennard-Jones potential, as well as Coulomb interactions that were summed up within the cut-off distance of 1 nm. Under these simulation conditions water molecules have been confirmed to form a massive aggregate on a hydrophobic surface and to spread over a hydrophilic surface,[31] which validates the employed conditions for the MD simulations.

Figure 5 shows a simulated diffusion process of 200 water molecules placed on a stack of bilayer rubrene SC/5-layer Au(100)/amorphous $SiO_2$ at 300 K. (Multimedia view) Among them, the velocity of Au, Si, and O atoms were set to zero, indicating that the Au and $SiO_2$ layers were fixed during the simulations. As depicted in Fig. 5(a), the water molecules were initially placed as a cube. As shown in the snapshots at 20 ps, Fig. 5(b), and 40 ps, Fig. 5(c), the centroid of the water molecules moved toward the rubrene surface and the water molecules spread to some extent over the surface. From these results it can be understood that (i) diffusion of water molecules was properly expressed by the MD simulations and (ii) water molecules were found to show a (partially) wetting behavior on rubrene. Importantly, no penetration of water molecules into the rubrene SC was observed within the MD simulations, which suggests that inter-molecular spacings are not an efficient pathway for water permeation. It should be noted here that we employed the rigid model and thus the intra-molecular degree of freedom was not allowed during the MD simulation. This was because inclusion of the intra-molecular motion resulted in disruption of the rubrene SC structure. There remains a possibility that proper inclusion of the intra-molecular motion leads to water permeation into the rubrene SC to some extent. However, considering the relatively wide dispersion of the HOMO-derived band of a rubrene SC,[34] it is reasonable to conclude that the inter-molecular interaction is strong enough to hinder the water permeation.

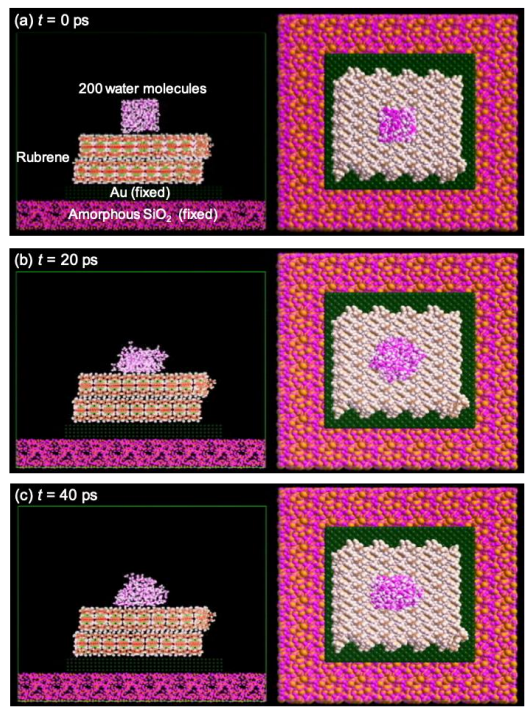

**Fig. 5.** MD simulation of water diffusion on a rubrene/Au/amorphous-SiO$_2$ stack at (a) 0 ps, (b) 20 ps, and (c) 40 ps. Left: side view. Right: top view. (Multimedia view)

Finally, we will discuss a possible form of the dominant pathway, i.e., a space at the SC/substrate interface. The present study employed a bottom-contact configuration for the fabrication of FETs. This configuration is superior to a top-contact one in terms of structural invariability upon electrode contact formation. This is because the top contact formation is known to form an ill-defined interface as a result of impingement of energetic metallic clusters to the soft organic layer during vacuum deposition.[35,36] In the bottom-contact configuration the pre-defined electrodes protrude from the substrate surface. If the SC lamination is not sufficiently conformal, a space should be formed around the electrode edges. In the present study, the total thickness of the electrodes was as small as 15 nm. The percolation threshold thickness for Au films on insulating substrates was determined to be around 7 nm.[37] Considering the evolution of the film structure from island, to mesh, to a continuous film, the thickness required to obtain a continuous electrode film can be estimated to be ~10 nm. Since the thickness of the Au electrodes in the present study is only ~5 nm above the required thickness for continuous films, the degree of the protrusion was minimized in the present devices. Even with such thin electrodes, formation of the space was found to be significant for SCFETs with a thick SC, as shown in Figs. 2(a–c). Ultrathin conductors[38] and embedded electrode structures[39] would be efficient for elimination of the space around the electrode edges.

**CONCLUSION**

In conclusion, we have investigated a water permeation pathway in laminated organic SC devices by utilizing the electrode-related instabilities as a probe of water permeation. By employing a rubrene SC as a channel of organic FETs, defects and grain boundaries were excluded as a possible pathway. The electrode-related instabilities, which are observable only in the presence of water in the surrounding environment, were found to be suppressed with thinner SCs. Thus, inter-molecular spacings can also be excluded as a possible pathway because the permeation via inter-molecular spacings should be more efficient with a thinner

SC. This finding was corroborated by MD simulations showing that penetration of water molecules via diffusion into single crystalline rubrene layers was found to be negligible within our simulation conditions. These results indicate that a space formed at the interface between the SC and the underlying substrate is a dominant pathway of water permeation in laminated organic SC devices. Therefore, the present study clearly shows the importance of conformality of the SC lamination onto the underlying substrate, which should be taken care generally in van der Waals-bonded systems including transferred two-dimensional crystals.[40,41] Furthermore, it is suggested that the electrode-related instabilities can be exploited as a tool to check how perfect the lamination is in van der Waals-bonded systems, as well as how defective the channel layer is in polycrystalline systems.

## ACKNOWLEDGMENTS

This work was supported by JSPS KAKENHI Grant Numbers JP17H01040, JP19H02561, and JP18K05253; and JST, PRESTO Grant Number JPMJPR17S6.

## DATA AVAILAVILITY STATEMENT

The data that support the findings of this study are available from the corresponding author upon reasonable request.

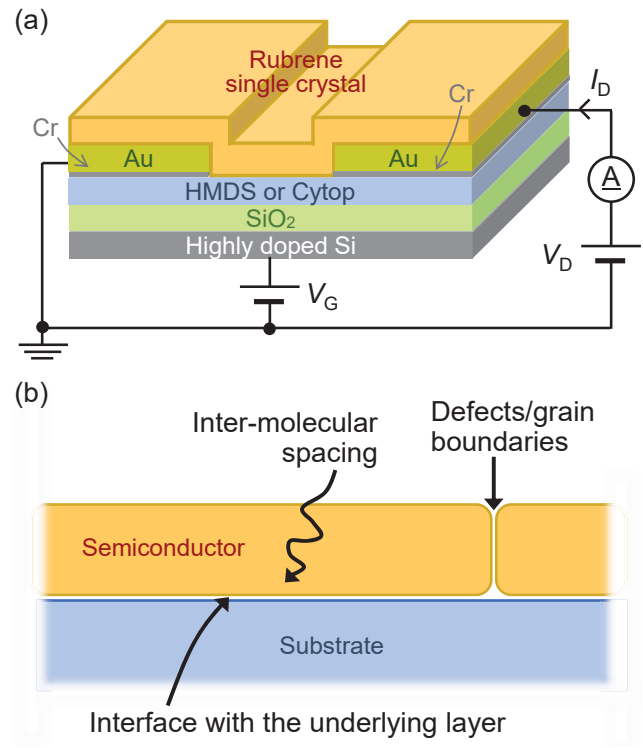

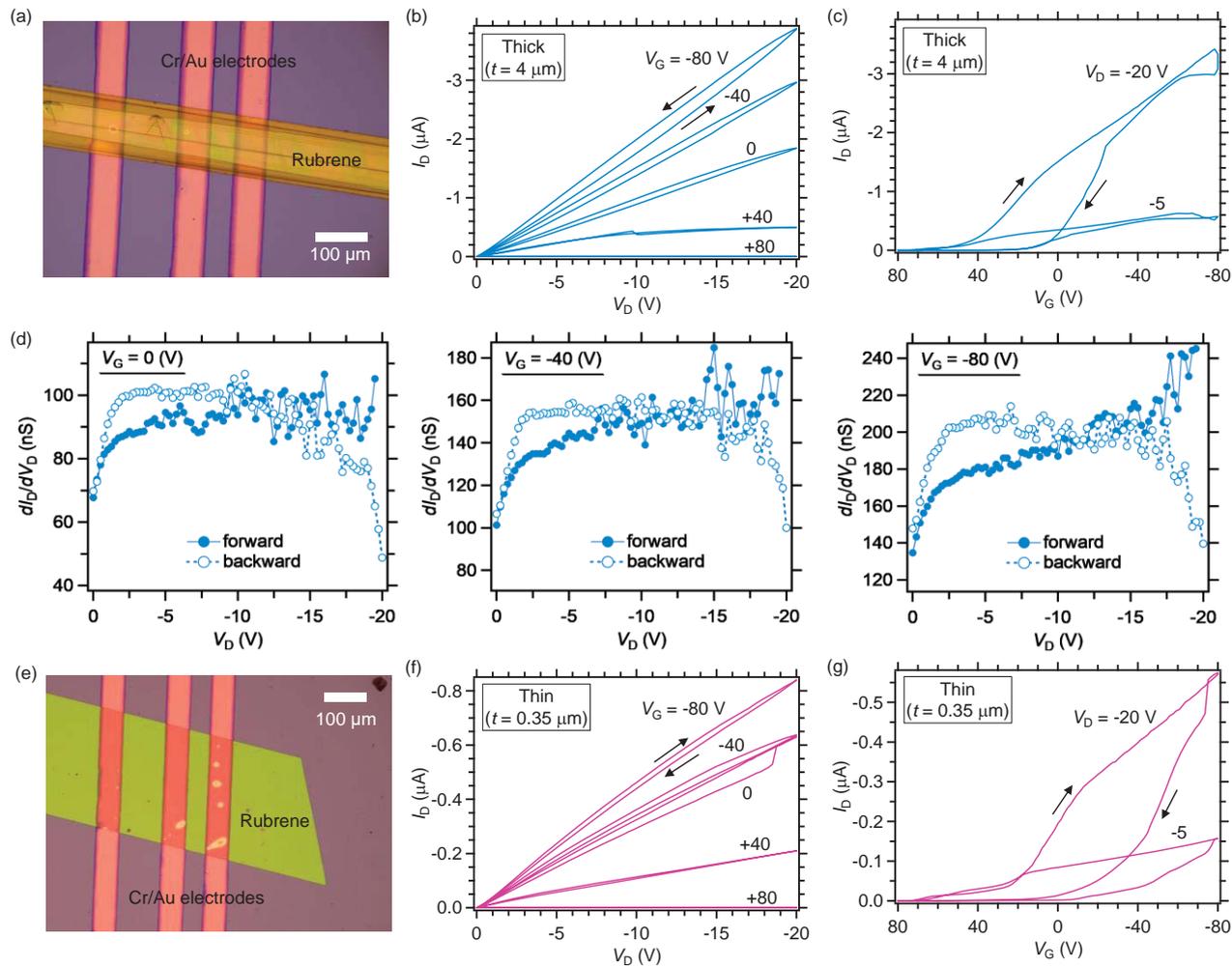

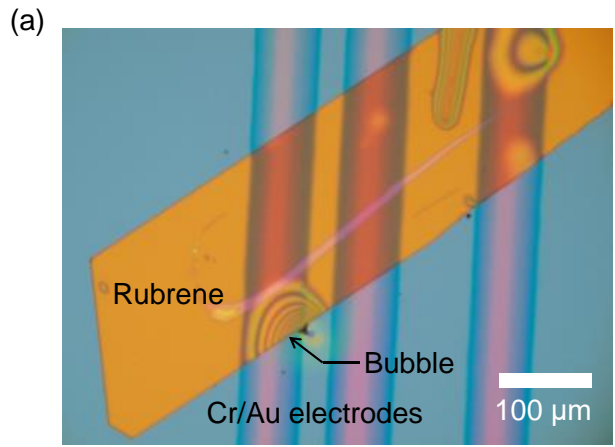 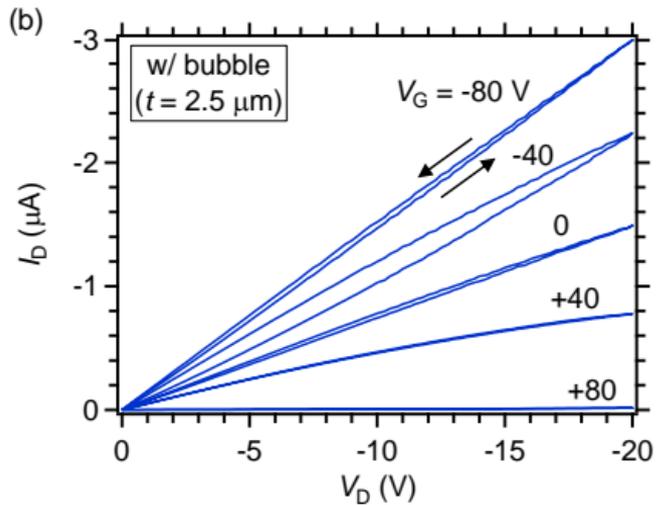 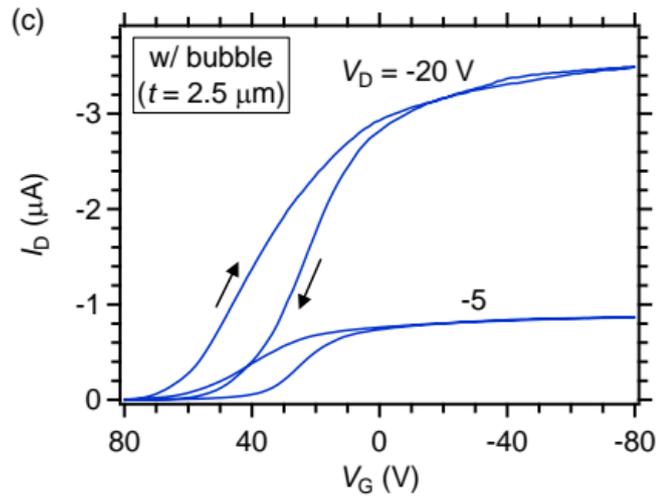
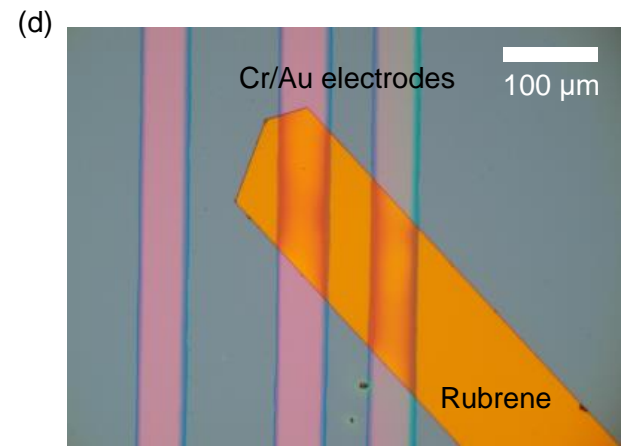 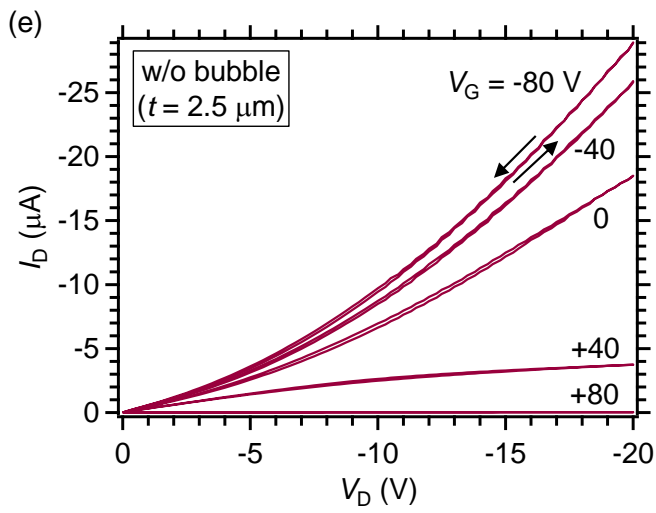 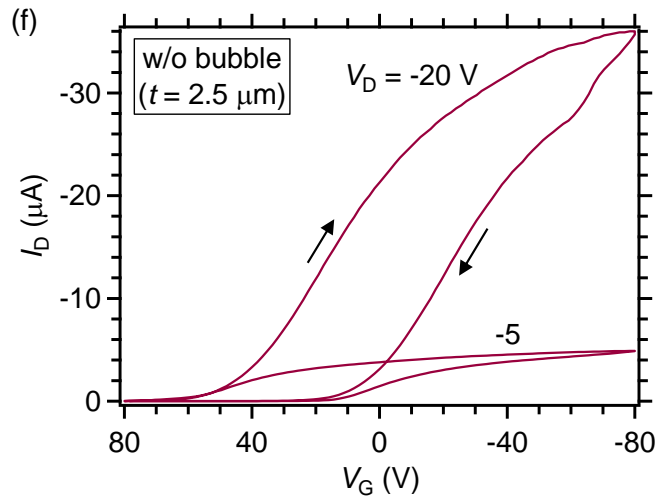

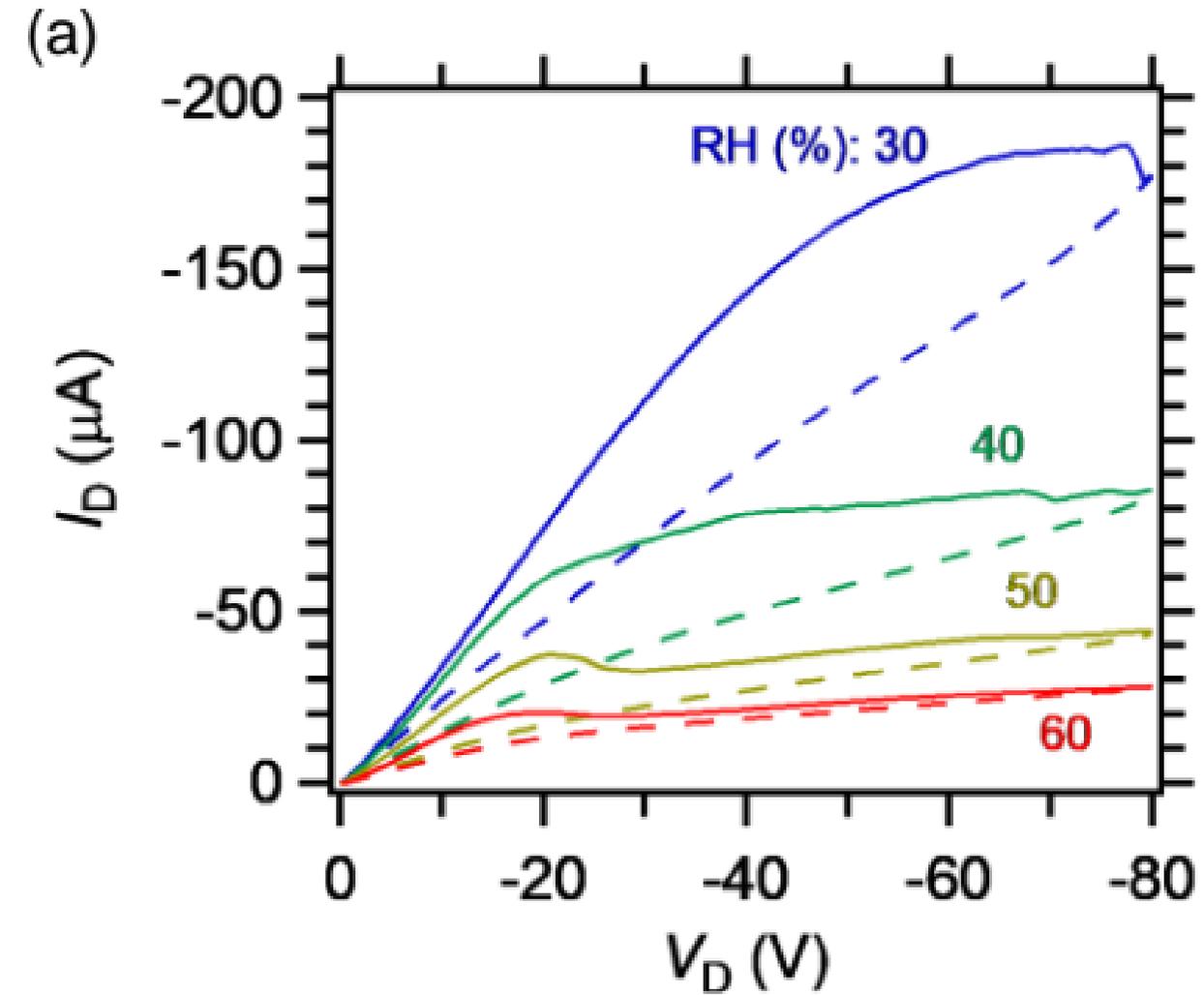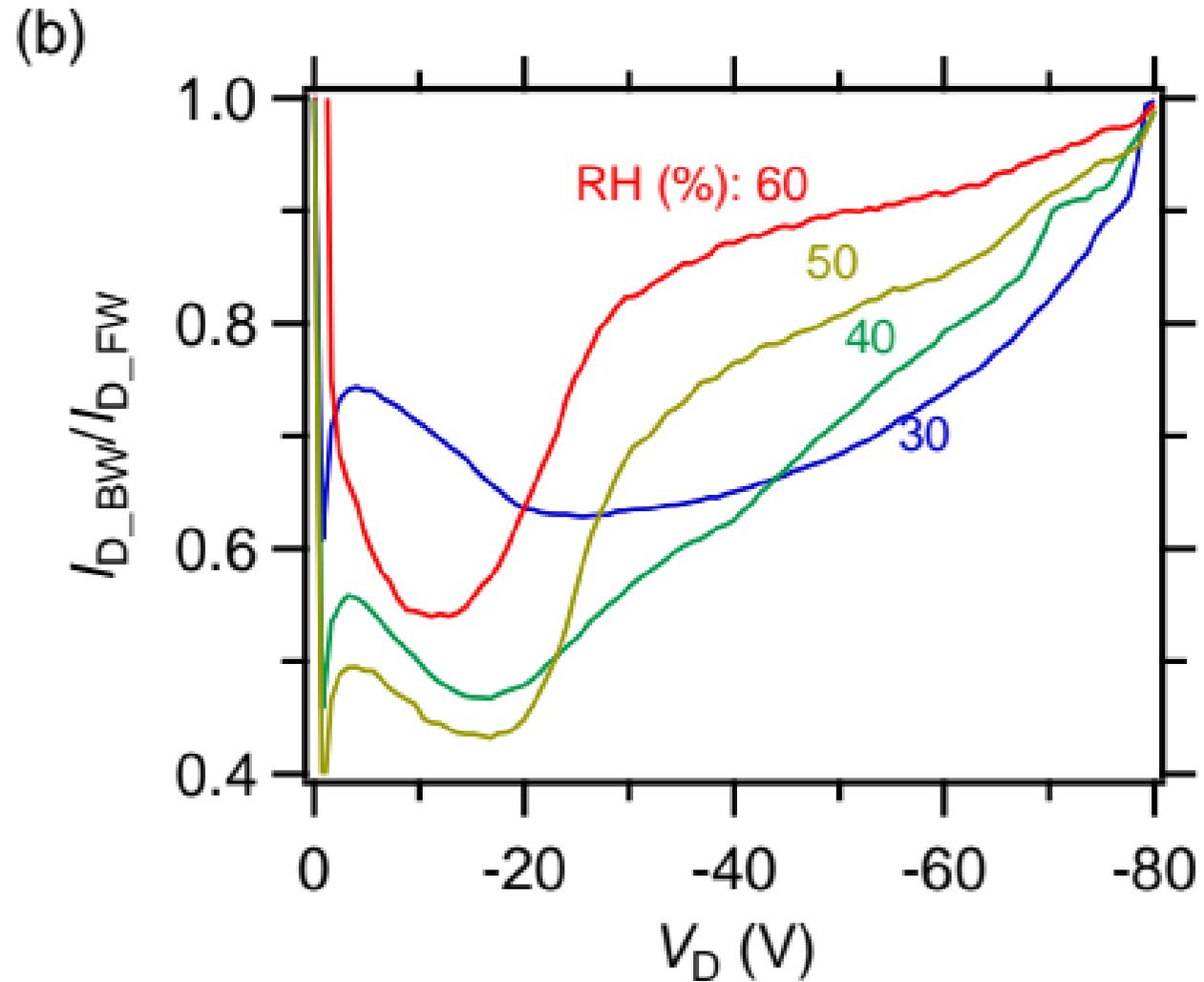

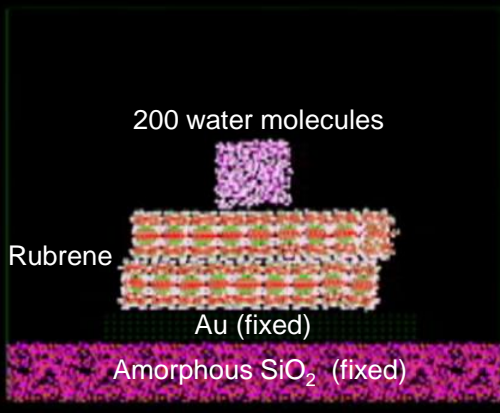 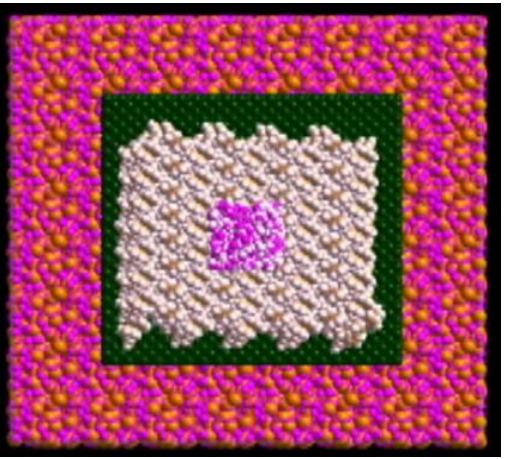

(a) $t = 0$ ps
200 water molecules
Rubrene
Au (fixed)
Amorphous SiO$_2$ (fixed)

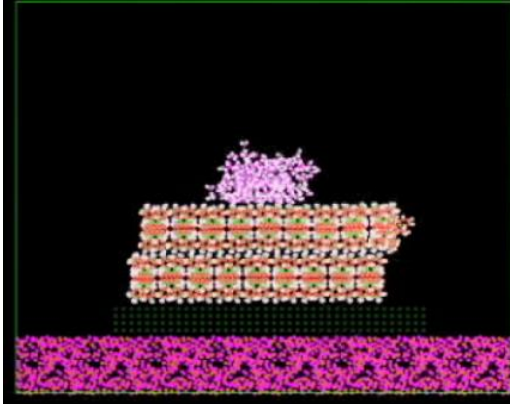 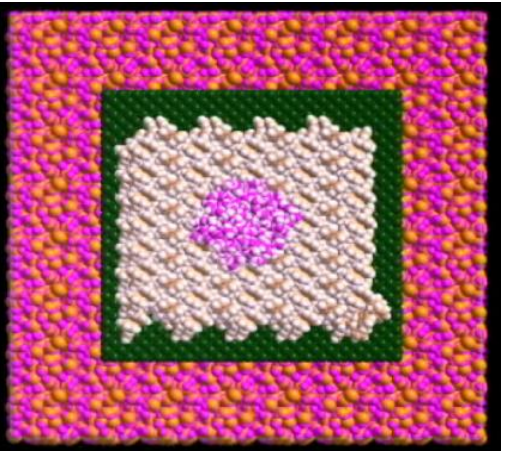

(b) $t = 20$ ps

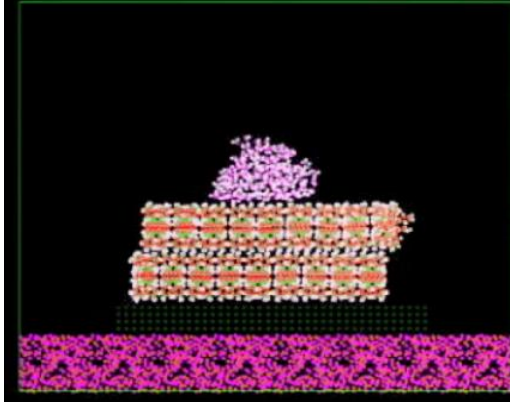 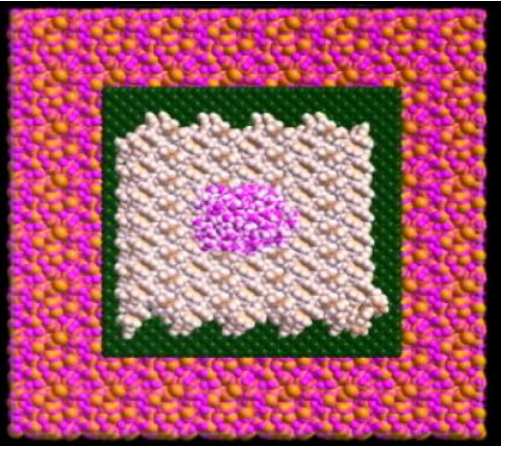

(c) $t = 40$ ps